\documentclass[aps,pra,twocolumn,10pt,superscriptaddress,showpacs,amsmath,amssymb,nofootinbib]{revtex4-1}

\usepackage{graphicx}
\usepackage{dcolumn}
\usepackage{multirow}

\newcommand{\SNR}{\text{SNR}}

\newcommand{\AEE}{A_{E1\text{-}E1}}
\newcommand{\AW}{A_{\text{W}}}
\newcommand{\AS}{A_{\text{S}}}

\newcommand{\APC}{A_{\text{PC}}}

\newcommand{\HW}{H_{\text{NSD}}}
\newcommand{\HS}{H_{\text{S}}}
\newcommand{\chiS}{\chi_{\text{S}}}
\newcommand{\chiW}{\chi_{\text{W}}}
\newcommand{\bra}[1]{\langle #1|}
\newcommand{\ket}[1]{| #1\rangle}
\newcommand{\CG}[3]{\langle #1;#2|#3\rangle}
\newcommand{\sixJ}[6]{\left\{
\begin{array}{ccc} #1 & #2 & #3 \\
#4 & #5 & #6 \end{array}\right\} }
\newcommand{\evec}{\boldsymbol{\epsilon}}

\newcommand{\kvec}{\mathbf{k}}
\newcommand{\dvec}{\mathbf{d}}
\newcommand{\muvec}{\boldsymbol{\mu}}
\newcommand{\Efield}{\mathbf{E}}
\newcommand{\Bfield}{\mathbf{B}}

\begin{document}
\title{Parity violation in two-photon $J=0\rightarrow 1$ transitions: Analysis of systematic errors}
\author{D.~R.~Dounas-Frazer}
\email{drdf@berkeley.edu}
\author{K.~Tsigutkin}
\author{D.~English}
\affiliation{Department of Physics, University of California at
Berkeley, Berkeley, California 94720-7300, USA}
\author{D.~Budker}
\affiliation{Department of Physics, University of California at
Berkeley, Berkeley, California 94720-7300, USA} \affiliation{Nuclear
Science Division, Lawrence Berkeley National Laboratory, Berkeley,
California 94720, USA}
\date{\today}


\begin{abstract}
We present an analysis of systematic sources of uncertainty in a recently proposed scheme for measurement of nuclear-spin-dependent atomic parity violation using two-photon $J=0\rightarrow 1$ transitions driven by collinear photons of the same frequency in the presence of a static magnetic field. Two important sources of uncertainty are considered: misalignment of applied fields, and stray electric and magnetic fields. The parity-violating signal can be discriminated from systematic effects using a combination of field reversals and analysis of the Zeeman structure of the transition.
\end{abstract}

\maketitle

%
%
%
%
%
%
%
\section{Introduction}
We previously proposed a method for measuring nuclear-spin-independent (NSI) atomic parity violation (APV) effects using two-photon transitions between states with zero total electronic angular momentum~\cite{DounasFrazer2011}. Recently we proposed another two-photon method, one that allows for NSI-background-free measurements of nuclear-spin-dependent (NSD) APV effects~\cite{DounasFrazerPAVI2011}, such as the nuclear anapole moment~\cite{Haxton2002}. The latter method, the \emph{degenerate photon scheme (DPS)}, exploits Bose-Einstein statistics (BES) selection rules for $J=0\rightarrow 1$ transitions driven by two collinear  photons of the same frequency~\cite{DeMille2000}. The general idea for the DPS was described in Ref.~\cite{DounasFrazerPAVI2011}. The present work complements Ref.~\cite{DounasFrazerPAVI2011} in the following ways: we derive expressions for amplitudes of allowed $E1$-$E2$ and $E1$-$M1$ transitions, and for $E1$-$E1$ transitions induced by the weak interaction and Stark effect; we analyze several major sources of systematic uncertainty  affecting the DPS; and, whereas Ref.~\cite{DounasFrazerPAVI2011} focused on mixing of the final state with nearby states of total electronic angular momentum $J=0$, we extend the analysis to include mixing with $J=2$ states as well.

%
%
%
%
%
%
%
\section{Degenerate Photon Scheme}
The proposed method uses two-photon transitions from an initial state of total electronic angular momentum $J_i=0$ to an opposite-parity $J_f=1$ final state (or \emph{vice versa}). The APV signal is due to interference of parity-conserving electric-dipole-electric-quadrupole ($E1$-$E2$) and electric-dipole-magnetic-dipole ($E1$-$M1$) transitions with parity-violating $E1$-$E1$ transitions induced by the weak interaction. This scheme is different from other multi-photon APV schemes~\cite{DounasFrazer2011,GunawardenaPRL2007,Guena2003,Cronin1998} in that the transitions are driven by collinear photons of the same frequency, and hence are subject to a Bose-Einstein statistics (BES) selection rule that forbids $E1$-$E1$ $J=0\rightarrow 1$ transitions~\cite{DeMille2000, English2010}. However, such transitions may be induced by perturbations that cause the final state to mix with opposite-parity $J\neq 1$ states, such as the NSD weak interaction and, in the presence of an external static electric field, the Stark effect. Because the NSI weak interaction only leads to mixing of the final state with other $J=1$ states, it cannot induce $J=0\rightarrow 1$ transitions. Thus NSI-background-free measurements of NSD APV can be achieved by exploiting two-photon BES selection rules.

Consider atoms illuminated by light in the presence of a static magnetic field $\Bfield$.
The optical field is characterized by polarization $\evec$, propagation  vector $\kvec$, frequency $\omega$, and intensity $\mathcal{I}$. Because circularly polarized light cannot excite a $J=0\rightarrow 1$ two-photon transition due to conservation of angular momentum, we assume that the light is linearly polarized. We choose the frequency to be half the energy interval  $\omega_{fi}$ between the ground state $\ket{i}$ and an excited state $\ket{f}$ of opposite  nominal parity. We work in atomic units: $\hbar=|e|=m_e=1$. The transition rate is~\cite{Faisal1987}:
\begin{equation}\label{eq:R}
R = (2\pi)^3 \alpha^2 \mathcal{I}^2 |A|^2 \frac{2}{\pi\Gamma},
\end{equation}
where $\alpha$ is the fine structure constant, and $A$ and $\Gamma$ are the amplitude and width of the transition. Energy eigenstates are represented as $\ket{i}=\ket{J_i I  F_i M_i}$, and likewise for $\ket{f}$. Here $J_i$, $I$, and $F_i$ are quantum numbers associated with the electronic, nuclear, and total angular momentum, respectively, and $M_i\in\{\pm F_i,\pm(F_i-1),\hdots\}$ is the projection of $F_i$ along the quantization axis ($z$-axis), which we choose along $\Bfield$.

\begin{figure}
\includegraphics[width=\columnwidth]{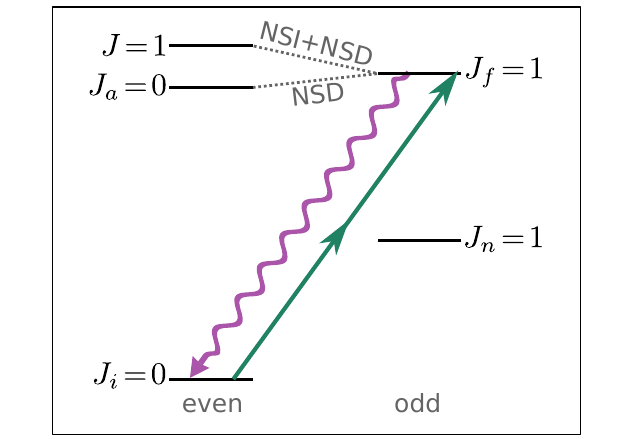}
\caption{\label{fig:1a} Energy level diagram.  Dotted lines indicate APV mixing of opposite-parity states, and upward- and downward-pointing arrows represent two-photon absorption and one-photon fluorescence, respectively.}
\end{figure}

The transition is enhanced by the presence of an intermediate state $\ket{n}$ of total electronic angular momentum $J_n=1$ whose energy lies about halfway between the energies of the initial and final states~(Fig.~\ref{fig:1a}). For typical situations, the energy defect $\Delta = \omega_{ni}-\omega_{fi}/2$ is large compared to the Rabi frequency $\Omega_\text{R}$ associated with the one-photon resonance involving the intermediate state. We assume that the scattering rate from $\ket{n}$ to $\ket{i}$ is small compared to the natural width $\Gamma_f$ of $\ket{f}$: $(\Omega_\text{R}/\Delta)^2\Gamma_n \ll \Gamma_f$. In this case, the system reduces to a two-level system consisting of initial and final states coupled by an effective optical field.

The parity-violating $E1$-$E1$ transition is induced by mixing of the final state with opposite-parity states via the weak interaction. In general, $\ket{f}$ may mix with states of electronic angular momentum $J = 0,  1$, or 2 according to the selection rules for NSD APV mixing~\cite{Khriplovich1991}. Mixing of the final state with  $J=1$ states results in a perturbed final state with electronic angular momentum~1 that cannot be excited via degenerate two-photon transitions. We assume mixing is dominated by a single state $\ket{a}$ of total angular momentum $J_a$, and consider the cases $J_a=0$ and $J_a=2$ separately.

\subsection{NSD APV mixing of $J=1$ and $J=0$ states}
When $\ket{f}$ mixes with a nearby $J_a=0$ state, only transitions for which $F_f=I$ may be induced by the weak interaction. Transitions to hyperfine levels $F_f=I\pm1$ that arise due to parity-conserving processes can be used as  APV-free references, important for discriminating APV from systematic effects. The amplitude for a degenerate two-photon $J=0\rightarrow 1$ transition is~
(Appendix~\ref{app:A}):
\begin{equation}\label{eq:A}
A = \APC + \AW,
\end{equation}
where
\begin{equation}\label{eq:APC}
\begin{split}
\APC = i\mathcal{Q}k_{-q}(\evec\cdot\evec)(-1)^q\CG{F_iM_i}{1q}{F_fM_f},
\end{split}
\end{equation}
and
\begin{equation}\label{eq:AW}
\begin{split}
\AW &= i\zeta_0(\evec\cdot\evec)\delta_{F_fF_i}\delta_{M_fM_i},
\end{split}
\end{equation}
are the amplitudes of the parity-conserving and weak-interaction-induced parity-violating transitions, respectively. Here $q=M_f-M_i$ is a spherical index, $k_q$ is the $q$th spherical component of $\hat{\kvec}$, $\langle F_iM_i;1q |F_fM_f\rangle$ is a Clebsch-Gordan coefficient, and $\delta_{F_fF_i}$ is the Kronecker delta. The quantities $\mathcal{Q}$ and $\zeta_0$ are
\begin{equation}\label{eq:Q}
\mathcal{Q} = \frac{Q_{fn}d_{ni}}{2\sqrt{15}\Delta} + \frac{\mu_{fn}d_{ni}}{3\sqrt{2}\Delta},
\end{equation}
and
\begin{equation}\label{eq:zeta0}
\zeta_0 =
\frac{\Omega_{fa}d_{an}d_{ni}}{3\omega_{fa}\Delta},
\end{equation}
where the reduced matrix elements $Q_{fn}=(J_f||\mathbb{Q}||J_n)$, $\mu_{fn}=(J_f||\mu||J_n)$, and $d_{ni}=(J_n||d||J_i)$ of the  electric quadrupole, magnetic dipole, and electric dipole moments, respectively, are independent of $F_f$ and $I$. Here  $\omega_{fa}=\omega_f-\omega_a$ is the energy difference of states $\ket{f}$ and $\ket{a}$, and $\Omega_{fa}$ is related to the matrix element of the NSD APV Hamiltonian $\HW$ by $\bra{f}\HW\ket{a}=i\Omega_{fa}$. The parameter $\Omega_{fa}$ must be a purely real quantity to preserve time reversal invariance~\cite{Khriplovich1991}.  Note that $\evec\cdot\evec=1$ for linear polarization, whereas $\evec\cdot\evec=0$ and hence $A=0$ for circular polarization, consistent with conservation of angular momentum. Hereafter, we assume $\evec\cdot\evec=1$.

\begin{figure}
\includegraphics[width=\columnwidth]{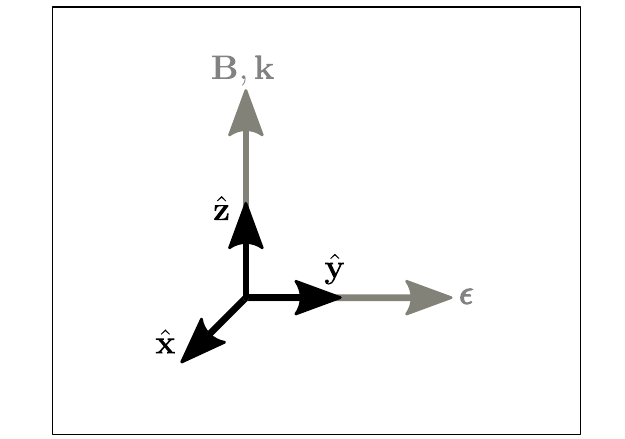}
\caption{\label{fig:1b} Field geometry. The propagation vector $\mathbf{k}$ may alternatively be anti-aligned with the magnetic field $\Bfield$.}
\end{figure}

The goal of the DPS is to observe interference of parity- violating and conserving amplitudes in the rate  $R$. When $M_f=M_i$, $R$ consists of a large parity conserving term proportional to $\mathcal{Q}^2$, a small parity violating term (the interference term) proportional to $\mathcal{Q}\zeta_0$, and a negligibly small term on the order of $\zeta_0^2$. The interference term is proportional to a pseudoscalar quantity that depends only on the field geometry, the \emph{rotational invariant}:
\begin{equation}\label{eq:rotinvar0}
\kvec\cdot\Bfield.
\end{equation}
The form of the rotational invariant follows from the fact that only $k_0\propto \kvec\cdot\Bfield$ contributes to the amplitude in Eq.~(\ref{eq:APC}) when $M_f=M_i$. Thus the interference term vanishes if $\Bfield$ and $\kvec$ are orthogonal. One way to achieve a nonzero rotational invariant is to orient $\kvec$ along $\Bfield$ (Fig.~\ref{fig:1b}).

We calculate the transition rate when $\Bfield$ is sufficiently strong to resolve magnetic sublevels of the final state, but not those of the initial state. This regime is realistic since Zeeman splitting of the initial and final states are proportional to the nuclear and Bohr magnetons, respectively. In this case, the total rate is the sum of rates from all magnetic sublevels of the initial state:
\begin{equation}
R\rightarrow \sum_{M_i}R(M_i).
\end{equation}
When the fields are aligned as in Fig.~(\ref{fig:1b}), the transition rate is
\begin{equation}\label{eq:A^2}
R_{\pm} \propto \frac{\mathcal{Q}^2M_f^2}{I(I+1)}\pm \frac{2\zeta_0\mathcal{Q}M_f}{\sqrt{I(I+1)}},
\end{equation}
where the positive (negative) sign is taken when $\kvec$ and $\Bfield$ are aligned  (anti-aligned), and we have omitted the term proportional to $\zeta_0^2$.

Reversals of applied fields are a powerful tool for discriminating APV from systematic effects. The interference term in~(\ref{eq:A^2}) changes sign when the relative alignment of $\kvec$ and $\Bfield$ is reversed, or when $M_f\rightarrow -M_f$. The \emph{asymmetry} is obtained by dividing the difference of rates upon a reversal by their sum:
\begin{equation}\label{eq:asymmetry}
\frac{R_+-R_-}{R_++R_-} = \frac{2\sqrt{I(I+1)}}{M_f}\frac{\zeta_0}{\mathcal{Q}},
\end{equation}
which is maximal when $M_f$ is small but nonzero. Reversals are sufficient to distinguish APV from many systematic uncertainties. Nevertheless,  there still exist systematic effects that give rise to \emph{spurious asymmetries}, which may mask APV.

We consider two potential sources of spurious asymmetry: misalignment of applied fields, and stray electric and magnetic fields. A stray electric field $\Efield$ may induce $E1$-$E1$ transitions via the Stark effect~\cite{Bouchiat1974,Bouchiat1975}. The amplitude of Stark-induced $J=0\rightarrow 1$ transitions is (Appendix~\ref{app:A}):
\begin{equation}\label{eq:AS}
\AS = \xi_0 E_{-q}(-1)^q\CG{F_iM_i}{1q}{F_fM_f},
\end{equation}
where
\begin{equation}\label{eq:xi0}
\xi_0 = \frac{d_{fa}d_{an}d_{ni}}{3\sqrt{3}\omega_{af}\Delta}.
\end{equation}
When $\kvec$ and $\Bfield$ are misaligned ($\kvec\times\Bfield\neq \mathbf{0}$), Stark-induced transitions may interfere with the allowed transitions yielding a spurious asymmetry characterized by the following rotational invariant:
\begin{equation}
(\Efield\times\kvec)\cdot\Bfield\equiv (\kvec\times\Bfield)\cdot\Efield.
\end{equation}
The resulting Stark-induced asymmetry is
\begin{equation}\label{eq:thetaE}
\frac{1}{M_f}\frac{\theta \xi_0 E}{\mathcal{Q}},
\end{equation}
where $\theta=|\hat{\kvec}\times\hat{\Bfield}|$ is the angle between the nominally collinear vectors $\kvec$ and $\Bfield$, and $E$ is defined by $\theta E \equiv (\kvec\times\Bfield)\cdot\Efield$. The spurious asymmetry (\ref{eq:thetaE}) may mask the APV asymmetry (\ref{eq:asymmetry}) because both exhibit the same behavior under field reversals.
However, because the Stark-induced transition amplitude is nonzero when $F_f \neq I$, APV and Stark-induced asymmetries can be determined unambiguously by comparing transitions to different hyperfine levels of the final state.

We propose to measure the transition rate by observing fluorescence of the excited, and assume that the transition is not saturated:
\begin{equation}\label{eq:Isat}
\mathcal{I}< \mathcal{I}_\text{sat}\equiv \Gamma/(4\pi\alpha\mathcal{Q}),
\end{equation}
where the saturation intensity $\mathcal{I}_\text{sat}$ is chosen so that $R=\Gamma$ when $\mathcal{I}=\mathcal{I}_\text{sat}$.  In this regime, fluorescence is proportional to the transition rate. The statistical sensitivity of this detection scheme is determined as follows: The number of excited atoms is
\begin{equation}
N_f=N_iR_{\pm}t \equiv N \pm N',
\end{equation}
where $N_i$ is the number of illuminated atoms, $t$ is the measurement time, and $N$ and $N'\ll N$ are the number of excited atoms due to parity-conserving and parity-violating processes. The signal-to-noise ratio is $\SNR=N'/\sqrt{N}$, or
\begin{align}\label{eq:SNR}
\SNR &= 8\pi\alpha \mathcal{I}\zeta_0\sqrt{N_it/\Gamma}\nonumber\\
&= 2(\mathcal{I}/\mathcal{I}_\text{sat})(\zeta_0/\mathcal{Q})\sqrt{N_i\Gamma t}.
\end{align}
The SNR is optimized by illuminating a large number of atoms with light that is intense, but does not saturate the $i\rightarrow f$ transition. Although purely statistical shot-noise dominated SNR does not depend on $\mathcal{Q}$, this parameter is still important in practice due to condition (\ref{eq:Isat}). Allowed $E1$-$E2$ and $E1$-$M1$ transitions are characterized by large $\mathcal{Q}$, which leads to small APV asymmetry. In the opposite case of forbidden $E1$-$E2$ and $E1$-$M1$ transitions (small $\mathcal{Q}$), an observable signal requires high light intensities, which may pose a technical challenge.

\subsection{NSD APV mixing of $J=1$ and $J=2$ states}
Mixing of $\ket{f}$ with nearby $J_a=2$ states is qualitatively similar to the previous case. Here we make the comparison explicit. The amplitude of the transition induced by NSD APV mixing of $\ket{f}$ and $\ket{a}$ is (Appendix~\ref{app:A}):
\begin{equation}\label{eq:AW2}
\AW = i\zeta_2\{\evec\otimes\evec\}_{2,-q}(-1)^q\CG{F_iM_i}{2q}{F_fM_f},
\end{equation}
where
\begin{equation}
\zeta_2 = \frac{\Omega_{fa}d_{an}d_{ni}}{\sqrt{15}\omega_{af}\Delta},
\end{equation}
and $\{\evec\otimes\evec\}_{2q}$ is the $q$th spherical component of the rank-2 tensor formed by taking the dyadic product of $\evec$ with itself~\cite{Varshalovich1988}.
For the geometry in Fig.~\ref{fig:1b}, the transition rate is
\begin{equation}\label{eq:R2}
R_{\pm} \propto C_1^2\mathcal{Q}^2\pm\sqrt{2/3}C_1C_2\zeta_2\mathcal{Q},
\end{equation}
where the positive (negative) sign is taken when $\kvec$ and $\Bfield$ are aligned  (anti-aligned), $C_k = \CG{IM_f}{k0}{F_fM_f}$ for $k=1,2$, and we have omitted a term proportional to $\zeta_2^2$. For simplicity, we focus on the case $F_f=I+1$ (the cases $F_f=I,I-1$ are similar). In this case, Eq.~(\ref{eq:R2}) becomes
\begin{equation}
R_{\pm} \propto \frac{F_f^2-M_f^2}{(I+1)(2I+1)}\left[\mathcal{Q}^2\pm \frac{2\zeta_2\mathcal{Q}M_f}{\sqrt{2I(I+2)}}\right],
\end{equation}
and the asymmetry is
\begin{equation}
\frac{R_+-R_-}{R_++R_-} = \frac{2M_f}{\sqrt{2I(I+2)}}\frac{\zeta_2}{Q},
\end{equation}
which is maximal when $M_f=I$. In the case of maximal asymmetry, the SNR is
\begin{align}\label{eq:SNR2}
\SNR &= 8\pi\alpha C_I \mathcal{I}\zeta_2\sqrt{N_i t/\Gamma}\nonumber\\
& =2C_I(\mathcal{I}/\mathcal{I}_\text{sat})(\zeta_2/\mathcal{Q})\sqrt{N_i\Gamma t}.
\end{align}
where $C_I=\sqrt{I/[2(I+1)(I+2)]}$ is a numerical coefficient and $\mathcal{I}_\text{sat}$ is given by Eq.~(\ref{eq:Isat}).

Static electric fields may induce a $J=0\rightarrow 1$ transition via Stark mixing of $\ket{f}$ and $\ket{a}$, giving rise to systematic effects that may mimic APV. When $J_a=2$, the amplitude of Stark-induced transitions is (Appendix~\ref{app:A}):
\begin{equation}\label{eq:AS2}
\AS = \xi_2[E_{-q}-3(\evec\cdot\Efield)\evec_{-q}](-1)^q\CG{F_iM_i}{1q}{F_fM_f},
\end{equation}
where
\begin{equation}\label{eq:xi2}
\xi_2 = \frac{d_{fa}d_{an}d_{ni}}{15\sqrt{3}\omega_{af}\Delta}.
\end{equation}
The spurious asymmetry due to Stark mixing is characterized by the rotational invariant
\begin{equation}
(\kvec\times\Bfield)\cdot[\Efield-3(\evec\cdot\Efield)\evec].
\end{equation}
Unlike for the $J_a=0$ case, both the Stark effect and the weak interaction may induce transitions to $F_f=I,I\pm1$ hyperfine levels of $\ket{f}$ when $J_a=2$, eliminating the possibility of using APV-free transitions to control systematic effects. However, the Stark- and weak-interaction- induced asymmetries have different dependence on $M_f$:
\begin{equation}
\frac{R_+-R_-}{R_++R_-} =\underbrace{\frac{2M_f}{\sqrt{2I(I+2)}}\frac{\zeta_2}{Q}}_{\text{APV}}
+\underbrace{\frac{(2I+1)M_f}{F_f^2-M_f^2}\frac{\theta \xi_2 \tilde{E}}{Q}}_{\text{Stark}},
\end{equation}
where $\tilde{E}$ is defined by $\theta\tilde{E}\equiv(\kvec\times\Bfield)\cdot[\Efield-3(\evec\cdot\Efield)\evec]$.
Thus APV can be distinguished from spurious asymmetries by analyzing the Zeeman structure of the transition, \emph{e.g.}, by comparing transitions to sublevels $M_f=I$ and $M_f=I-1$ of the final state.

As a final note, in addition to the rotational invariant (\ref{eq:rotinvar0}), there is a second parity-violating rotational invariant that arises when $\ket{f}$ mixes with $J=2$ states:
\begin{equation}
(\kvec\cdot\Bfield)(\evec\cdot\Bfield)^2.
\end{equation}
This rotational invariant describes APV interference in transitions for which $M_f=M_i\pm1$.

%
%
%
%
%
%
%
\section{Applications of DPS}\label{sec:application}

\begin{table}
\caption{\label{tab:1} Available atomic data for application of DPS to Sr and Ra. Here $a_0$ is the Bohr radius.}
\begin{ruledtabular}
\renewcommand{\arraystretch}{1.2}
\begin{tabular}{lrc}
 & \multicolumn{1}{c}{Transition} $(1\rightarrow 2)$ & $d_{21}/(ea_0)$ \\ \hline
 Sr\footnote{Ref.~\cite{Sansonetti2010}}
  & $5s^2\;{}^1S_0\rightarrow 5s5p\;{}^1P_1$ & 5.4 \\
 Ra\footnote{Ref.~\cite{Dzuba2006}}
  & $7s^2\;{}^1S_0\rightarrow 7s7p\;{}^1P_1$ & 5.8 \\
  & $7s7p\;{}^1P_1\rightarrow 7s6d\;{}^3D_2$ & 0.6\\
  & $7s6d\;{}^3D_2\rightarrow 7s7p\;{}^3P_1$ & 4.8 \\
\end{tabular}
\end{ruledtabular}
\end{table}

We now turn our attention to the two-photon 462 nm $5s^2\;{}^1S_0\rightarrow5s9p\;{}^1P_1$ transition in ${}^{87}$Sr ($Z=38$, $I=9/2$). The transition is enhanced by the intermediate $5s5p\;{}^1P_1$ state ($\Delta = 34$~cm$^{-1}$), and the parity-violating $E1$-$E1$ transition is induced by NSD APV mixing of the $5s9p\;{}^1P_1$ and $5s10s\;{}^1S_0$ states ($\omega_{af}=184$~cm$^{-1}$). We used expressions presented in Ref.~\cite{Dzuba_arXiv} to calculate the NSD APV matrix element: $\Omega_{fa}\approx 10\kappa$~s$^{-1}$, where $\kappa$ is a dimensionless constant of order unity that characterizes the strength of NSD APV. The width of the transition is determined by the natural width $\Gamma=1.15\times10^7$~s$^{-1}$ of the $5s9p\;{}^1P_1$ state~\cite{Sansonetti2010}. Other essential atomic parameters are given in Table~\ref{tab:1}. Resolution of the magnetic sublevels of the final state requires a magnetic field larger than $2\Gamma/g\approx10$~G, where $g\approx 0.1$ is the Land\'e factor of the $F=I$ hyperfine level of the $5s9p~^1P_1$ state. We estimate that $d_{an}\approx ea_0$, $\mathcal{Q}_{fn}/d_{an}\approx\alpha/2$, and $\mu_{fn}\ll Q_{fn}$.  Then the APV asymmetry associated with this system is about $4\kappa\times10^{-8}$.

Spurious asymmetries due to stray electric fields can be ignored when $\theta E\ll 2(2I+1)\sqrt{I(I+1)}(\zeta_0/\xi_0)\approx 2$~mV/cm. In ongoing APV experiments in Yb~\cite{Tsigutkin2010}, stray electric fields  on the
order of 1~V/cm have been observed. Assuming a similar magnitude of stray fields for the Sr system, spurious Stark-induced asymmetries can be ignored by controlling misalignment of the light propagation and the magnetic field to better than $\theta<0.1^\circ$. Regardless of misalignment errors, APV can be discriminated from Stark-induced asymmetries by comparing transitions to the $F\neq 9/2$ hyperfine levels of the final $5s9p\;{}^1P_1$ state.

To estimate the SNR, we consider experimental parameters similar to those of Ref.~\cite{Tsigutkin2009}: $N_i\approx 10^7$ atoms illuminated by a laser beam of characteristic radius $0.3$~mm.  Optimal statistical sensitivity is realized when $\mathcal{I} =\mathcal{I}_{\text{sat}}\approx6\times10^5$~W/cm$^2$. In this case, Eq.~(\ref{eq:SNR}) yields $\SNR\approx \kappa \times 10^{-3}\sqrt{t/\text{s}}$. The saturation intensity corresponds to light power of about 2~kW at 462~nm. High light powers may be achieved in a running-wave power buildup cavity.  With this level of sensitivity,  about 300~hours of measurement time are required to achieve unit SNR. The projected asymmetry and SNR for the Sr system are to their observed counterparts in the most precise measurements of NSD APV in Tl~\cite{Vetter1995}.


Another potential candidate for the DPS is the 741 nm $7s^2\;^1S_0\rightarrow 7s7p\;^3P_1$ transition in unstable ${}^{225}$Ra ($Z=88$, $I=3/2$, $t_{1/2}=15$~days).  This system lacks an intermediate state whose energy is nearly half that of the final state; the closest state is $7s7p\;{}^1P_1$ ($\Delta \approx 14000$~cm$^{-1}$). Nevertheless, it is a good candidate for the DPS, partly due to the presence of nearly-degenerate opposite-parity levels $7s7p\;^3P_1$ and $7s6d\;{}^3D_2$ ($\omega_{af}=5$~cm$^{-1}$).  In this system, NSD APV mixing arises due to nonzero admixture of configuration $7p^2$ in the $7s6d\;{}^3D_2$ state~\cite{FlambaumPRAR1999}. Numerical calculations yield $\Omega_{fa}d_{an}/\omega_{af}\approx2\kappa\times10^{-9}$~$ea_0$~\cite{Dzuba2000} and $\Gamma=2.8\times10^6$~s$^{-1}$~\cite{Dzuba2006}. Other essential atomic parameters are given in Table~\ref{tab:1}. Like for the Sr system, we estimate that $Q_{fn}/d_{an}=\alpha/2$ and $\mu_{fn}\ll Q_{fn}$, yielding an approximate asymmetry of $7\kappa\times10^{-6}$. Laser cooling and trapping of ${}^{225}$Ra has been demonstrated~\cite{Guest2007}, producing about $N_i\approx 20$ trapped  atoms. When $\mathcal{I}=\mathcal{I}_\text{sat}\approx10^{8}$~W/cm$^2$, Eq.~(\ref{eq:SNR2}) gives $\SNR = \kappa\times10^{-2}\sqrt{t/\text{s}}$.  For a laser beam of 0.3 mm, the saturation intensity corresponds to light power of about 300~kW at 741~nm. These estimates suggest that unit SNR can be realized in under 3~hours of observation time. Compared to the Sr system, the Ra system potentially exhibits both a much larger asymmetry and a much higher statistical sensitivity.

%
%
%
%
%
%
%
\section{Summary and discussion}
In conclusion, we presented a method for measuring NSD APV without NSI background. The proposed scheme uses two-photon $J=0\rightarrow 1$ transitions driven by collinear photons of the same frequency, for which NSI APV effects are suppressed by BES. We described the criteria necessary for optimal SNR and APV asymmetry, and identified transitions in ${}^{87}$Sr and ${}^{225}$Ra that are promising candidates for application of the DPS.

%
%
%
%
%
%
%
\acknowledgments The authors acknowledge helpful discussions with D.~P.~DeMille, V.~Dzuba, V.~Flambaum, M.~Kozlov, and N.~A.~Leefer. This work has been supported by NSF.

\appendix

\section{Derivation of transition amplitudes}\label{app:A}
In this appendix, we derive amplitudes for induced $E1$-$E1$
$J=0\rightarrow 1$ transitions between opposite parity states.

\subsection{Bose-Einstein statistics selection rules}
Here we provide a brief review of BES selection rules for $J=0\rightarrow 1$ transitions driven by degenerate ($\omega_1=\omega_2\equiv\omega$), co-propagating ($\kvec_1=\kvec_2\equiv\kvec$) photons~\cite{DeMille2000}. Since the only transitions of relevance are of this type, they are referred to as simply ``degenerate transitions" without cumbersome qualifiers. We ignore hyperfine interaction (HFI) effects by assuming that there is zero nuclear spin.

It must be possible to write the absorption amplitude $\mathcal{A}$ for a degenerate transition in terms of the only quantities available: the polarizations of the two photons, $\evec_1$ and $\evec_2$; the final polarization of the atom in its excited $J=1$ state, $\evec_e$; and the photon momentum $\kvec$. With the requirement of gauge invariance of the photons ($\evec_{1,2}\cdot\kvec=0$), only three forms of $\mathcal{A}$ are possible:
\begin{subequations}
\begin{align}
A_a &\propto (\evec_1\times\evec_2)\cdot\evec_e;\\
A_b &\propto (\evec_1\cdot\evec_2)(\evec_e\cdot\kvec);\\
A_c &\propto [(\evec_1\times\evec_2)\cdot\kvec](\evec_e\cdot\kvec).
\end{align}
\end{subequations}
Amplitudes $A_a$ and $A_c$ are odd under photon interchange, and hence vanish because photons obey BES. However, amplitude $A_b$ is even and may yield a nonzero absorption amplitude. In the case of degenerate transitions between atomic states of the same total parity, $A_b$ vanishes because it is odd under spatial inversion. Hence degenerate transitions between like-parity states are forbidden by BES selection rules\footnote{
Parity-conserving perturbations, such as the Zeeman effect or the hyperfine interaction, may induce degenerate transitions between like-parity states via two mechanisms: splitting of the intermediate state into non-degenerate sublevels, and mixing of the final state with nearby like-parity $J\neq 1$ states~\cite{English2009}.
}. However, degenerate transitions may be allowed when the initial and final states are of opposite parity.

When $\evec_1=\evec_2\equiv\evec$, as would be the case if the photons were absorbed from the same laser beam, the degenerate transition amplitude reduces to $A_b \propto (\evec\cdot\evec)\kvec\cdot\evec_e$. Therefore, the amplitude of a degenerate transition between opposite parity states is
\begin{equation}\label{eq:Ab}
A_b = i(\evec\cdot\evec)\mathcal{Q}k_{-M}(-1)^M,
\end{equation}
where $k_{-M}(-1)^M$ is the projection of $\hat{\kvec}$ onto the spin of the excited atom and the factor of $i$ ensures time reversal invariance.

\subsection{Wigner-Eckart theorem}

We use the following convention for the Wigner-Eckart theorem (WET).
Let $T_k$ be an irreducible tensor of rank $k$ with spherical
components $T_{kq}$ for $q\in\{0,\pm1,\hdots,\pm k\}$. Then the WET
is~\cite{Sobelman1992}
\begin{equation}\label{eq:WET}
\begin{split}
&\bra{J_2IF_2M_2}T_{kq}\ket{J_1IF_1M_1}=\\
&\quad=\frac{(J_2IF_2||T_k||J_1IF_1)}{\sqrt{2F_2+1}}\CG{F_1M_1}{kq}{F_2M_2},
\end{split}
\end{equation}
where $(J_2IF_2||T_k||J_1IF_1)$ is the reduced matrix element of
$T_k$ and $\CG{F_1M_1}{kq}{F_2M_2}$ is a Clebsch-Gordan coefficient.
If $T_k$ commutes with the nuclear spin $\mathbf{I}$, then its
reduced matrix element satisfies~\cite{Sobelman1992}
\begin{equation}\label{eq:sixJ}
\begin{split}
&\frac{(J_2IF_2||T_k||J_1IF_1)}{\sqrt{2F_2+1}}=(-1)^{J_2+I+F_1+k}(J_2||T_k||J_1)\times
\\ &\quad\times\sqrt{2F_1+1}
\sixJ{J_2}{F_2}{I}{F_1}{J_1}{k},
\end{split}
\end{equation}
where $(J_2||T_k||J_1)$ is the reduced matrix element of $T_k$ in
the decoupled basis, and the quantity in the curly braces is a $6j$
symbol.

\subsection{$E1$-$M1$ and $E1$-$E2$ transition amplitudes}
In the following, summation over the magnetic sublevels $M_n$ of the intermediate state is implied. The $E1$-$M1$ and $E1$-$E2$ transition amplitudes are
\begin{align}
A_{b_1} &= 2\bra{f} [(\kvec\times\evec)\cdot\muvec] \frac{\ket{n}\bra{n}}{\Delta} (\evec\cdot\dvec) \ket{i}\nonumber\\
&=i\left(\frac{2\mu_{fn}d_{ni}}{3\sqrt{2}\Delta}\right)k_{-M}(-1)^{M}\CG{F_iM_i}{1q}{F_fM_f},\label{eq:Ab1}
\end{align}
and
\begin{align}
A_{b_2} &= \bra{f} [i\{\kvec\otimes\evec\}_2\cdot \mathbb{Q}] \frac{\ket{n}\bra{n}}{\Delta} (\evec\cdot\dvec) \ket{i}\nonumber\\
&=i\left(\frac{Q_{fn}d_{ni}}{2\sqrt{15}\Delta}\right)k_{-M}(-1)^{M}\CG{F_iM_i}{1q}{F_fM_f},\label{eq:Ab2}
\end{align}
respectively. Here $\muvec$, $\dvec$, and $\mathbb{Q}$ are the magnetic dipole, electric dipole, and electric quadrupole moments of the atom.To derive Eqs.~(\ref{eq:Ab1}) and (\ref{eq:Ab2}), we have assumed $\evec\cdot\evec=1$, as is the case for linear polarization, and we have omitted a common factor of $(-1)^{I-F_f}$. Equations~(\ref{eq:APC}) and (\ref{eq:Q}) follow from the definition $\APC\equiv A_{b_1}+A_{b_2}$.

\subsection{Induced $E1$-$E1$ transitions}

$E1$-$E1$ transitions may be  induced by mixing of the states $\ket{f}$ and
$\ket{a}$ due to both the weak interaction and Stark effect. The
final state of the transition is the perturbed state
$\ket{f}+\chi^\ast\ket{a}$, where $\chi$ is a small dimensionless
parameter that depends on the details of the perturbing Hamiltonian.
The amplitude for the induced $E1$-$E1$ transition is~\cite{Faisal1987}
\begin{equation}\label{eq:AEE_start}
\begin{split}
\AEE = \chi\bra{a}\evec\cdot\dvec
\frac{\ket{n}\bra{n}}{\omega_{ni}-\omega}\evec\cdot\dvec\ket{i}.
\end{split}
\end{equation}
Here $\dvec$ is the electric-dipole moment of the atom and summation
over the hyperfine levels and magnetic sublevels of the states
$\ket{n}$ and $\ket{a}$ is implied.

In Eq~(\ref{eq:AEE_start}), the quantity $\bra{a}\cdots\ket{i}$ is
the amplitude of the allowed degenerate two-photon $i\rightarrow a$
transition. It can be expressed as the contraction of two
irreducible tensors:
\begin{equation}\label{eq:a...i}
\bra{a}\evec\cdot\dvec
\frac{\ket{n}\bra{n}}{\omega_{ni}-\omega}\evec\cdot\dvec\ket{i} =
\sum_{k,q} \{\evec\otimes\evec\}_{kq}^\ast\bra{a}T_{kq}\ket{i},
\end{equation}
where
\begin{equation}\label{eq:dyadic}
\{\evec\otimes\evec\}_{kq} =
\sum_{\mu,\nu}\CG{1\mu}{1\nu}{kq}\epsilon_\mu\epsilon_\nu,
\end{equation}
is the tensor of rank $k=0,2$ formed by the dyadic product of
$\evec$ with itself, and $T_{kq}$ is a tensor whose matrix elements
we wish to express in terms of those of the dipole moment $\dvec$.
Neglecting hyperfine splitting of the intermediate state, $T_k$
commutes with $\mathbf{I}$. Then, since $J_i=0$, we have
\begin{equation}
\bra{a}T_{kq}\ket{i} =
(-1)^{I-F_a+k}\frac{(J_a||T_k||J_i)}{\sqrt{2J_a+1}}\CG{F_iM_i}{kq}{F_fM_f},
\end{equation}
for $k=J_a$ and $q=M_a-M_i$. Using the WET to simplify the left-hand
side of Eq.~(\ref{eq:a...i}), we find
\begin{equation}
(J_a||T_k||J_i) = \frac{1}{\sqrt{3}}\frac{d_{an}d_{ni}}{\Delta},
\end{equation}
where $d_{fa} = (J_f||d||J_a)$ is the reduced matrix element of the
electric dipole operator. When $k\neq J_a$, the matrix element
$\bra{a}T_{kq}\ket{i}$ vanishes. Therefore, only the tensor
$\{\evec\otimes\evec\}_k$ of rank $k=J_a$ contributes to the
$i\rightarrow a$ transition. Note that the tensor of rank $k=1$
satisfies
$\{\evec\otimes\evec\}_{1q}\propto(\evec\times\evec)_q\equiv 0$, and
hence the $J_i=0\rightarrow J_a=1$ transition has zero amplitude,
consistent with more general selection rules for degenerate
two-photon transitions~\cite{DeMille2000}.

When the mixing of $\ket{f}$ and $\ket{a}$ is due to the weak
interaction alone, the perturbation parameter is given by
$\chi=\chiW$, where
\begin{equation}\label{eq:chiweak}
\chiW = \frac{\bra{f}\HW\ket{a}}{\omega_{fa}} \equiv \frac{i\Omega_{fa}}{\omega_{fa}},
\end{equation}
for $F_a=F_f$ and $M_a=M_f$. Equations
(\ref{eq:AW})and (\ref{eq:AW2}) follow from the definitions
$\zeta_k\equiv(\Omega_{fa}/\omega_{fa})(J_a||T_k||J_i)/\sqrt{2k+1}$ for $k=J_a=0,2$.

In the presence of a static electric field $\Efield$, the
perturbation parameter becomes $\chi = \chiW + \chiS$, where $\chiW$
is given by Eq.~(\ref{eq:chiweak}) and
\begin{align}\label{eq:HSfa}
\chiS &= \frac{\bra{f}\HS\ket{a}}{\omega_{fa}},
\end{align}
where $\HS=-\dvec\cdot\Efield$ is the Stark Hamiltonian. In this
case, $\AEE=\AW+\AS$, where $\AW\propto\chiW$ and $\AS\propto\chiS$
are the amplitudes of the transitions induced by the weak
interaction and Stark effect, respectively.

For a general $J=J_i\rightarrow J_f$ transition, the Stark-induced
$E1$-$E1$ amplitude may have contributions from each of the irreducible
tensors that can be formed by combining $\{\evec\otimes\evec\}_{0}\propto(\evec\cdot\evec)$ or $\{\evec\otimes\evec\}_{2}$
with $\Efield$. There are four such tensors: one each of ranks 2 and
3, and two of rank 1. However, for $J=0\rightarrow 1$ transition, only the rank-1
tensors contribute. These tensors are~\cite{Varshalovich1988}:
\begin{equation}\label{eq:T0}
\{\Efield\otimes\{\evec\otimes\evec\}_0\}_{1q} =
-\frac{1}{\sqrt{3}}(\evec\cdot\evec)E_q
\end{equation}
and
\begin{equation}\label{eq:T2}
\{\Efield\otimes\{\evec\otimes\evec\}_2\}_{1q} =
\sqrt{\frac{1}{15}}[E_q-3(\evec\cdot\Efield)\epsilon_q].
\end{equation}
Stark mixing of $\ket{f}$ with $\ket{a}$ gives rise to a Stark-induced amplitude whose dependence on applied fields is described by either the tensor in Eqs.~(\ref{eq:T0}) or the one in Eq.~(\ref{eq:T2}) depending on whether $J_a=0$ or $J_a=2$. The corresponding amplitudes are given by Eqs.~(\ref{eq:AS}) and (\ref{eq:AS2}), and the parameters $\xi_0$ and $\xi_2$ can be expressed in terms of the
reduced dipole matrix elements by applying the WET to
Eq.~(\ref{eq:AEE_start}) with $\chi=\chiS$. This procedure yields Eqs.~(\ref{eq:xi0}) and (\ref{eq:xi2}).

\end{document}